# Quantum oscillations in two coupled charge qubits

Yu. A. Pashkin,[†*] T. Yamamoto,[†‡] O. Astafiev,[†]

Y. Nakamura,[†‡] D. V. Averin[§] & J. S. Tsai[†‡]

[†] *The Institute of Physical and Chemical Research (RIKEN), Wako, Saitama 351-0198, Japan*

[‡] *NEC Fundamental Research Laboratories, Tsukuba, Ibaraki 305-8501, Japan*

[§] *Department of Physics and Astronomy, SUNY Stony Brook, Stony Brook, NY 11794-3800 USA*

**A practical quantum computer,[1] if built, would consist of a set of coupled two-level quantum systems (qubits). Among a variety of qubits implemented,[2] solid-state qubits are of particular interest because of their possible integration. In the Josephson-junction qubits,[3,4] coherence of Cooper pair tunnelling in the superconducting state is exploited.[5-10] Several types of such qubits have been implemented[5-8] demonstrating the potential of Josephson-junction circuits. Despite an apparent progress in implementing individual solid-state qubits, there have been no experimental reports so far on multi-bit gates required for building a real quantum computer. Here we report a new circuit comprising two coupled charge qubits. Using a pulse technique, we can coherently mix quantum states and observe quantum oscillations whose spectrum reflects interaction between the qubits. Our results demonstrate the feasibility of coupling of multiple solid-state qubits and indicate the existence of entangled two-qubit states.**

One of the physical realisations of a solid-state qubit is provided by a Cooper-pair box.[11] The two charge states of the box, say $|0\rangle$ and $|1\rangle$, differing by one Cooper pair are coherently mixed by the Josephson coupling as was confirmed experimentally.[12,13] Quantum state manipulation of such a system can be done by using a non-adiabatic pulse technique, and the read-out can be performed by a properly biased probe electrode.[5] Here we make one step further on the way to implementation of quantum logic gates by integrating two charge qubits and demonstrating their interaction.

The two charge qubits of our circuit are electrostatically coupled by an on-chip capacitor (Fig. 1). The right qubit has a SQUID geometry to allow the control of the Josephson coupling to its reservoir. Both qubits have a common pulse gate but separate dc gates, probes and reservoirs. The pulse gate has nominally equal coupling to each box. The Hamiltonian of the system in the two-qubit charge basis $|00\rangle$, $|10\rangle$, $|01\rangle$ and $|11\rangle$ reads:

$$H = \begin{bmatrix} E_{00} & -\frac{1}{2}E_{J1} & -\frac{1}{2}E_{J2} & 0 \\ -\frac{1}{2}E_{J1} & E_{10} & 0 & -\frac{1}{2}E_{J2} \\ -\frac{1}{2}E_{J2} & 0 & E_{01} & -\frac{1}{2}E_{J1} \\ 0 & -\frac{1}{2}E_{J2} & -\frac{1}{2}E_{J1} & E_{11} \end{bmatrix} \quad (1)$$

where $E_{n1n2} = E_{c1}(n_{g1}-n_1)^2 + E_{c2}(n_{g2}-n_2)^2 + E_m(n_{g1}-n_1)(n_{g2}-n_2)$ is the total electrostatic energy of the system ($n_1$, $n_2 = 0$, 1 is the number of excess Cooper pairs in the first and the second box), $E_{J1}$ ($E_{J2}$) is the Josephson coupling energy of the first (second) box and the reservoir, $E_{c1,2} = 4e^2 C_{\Sigma 2,1}/2(C_{\Sigma 1}C_{\Sigma 2} - C_m^2)$ are the effective Cooper-pair charging energies, $C_{\Sigma 1,2}$ are the sum of all capacitances connected to the corresponding island including the coupling capacitance $C_m$, $n_{g1,2} = (C_{g1,2}V_{g1,2} + C_p V_p)/2e$ are the normalised charges induced on the corresponding qubit by the dc and pulse gate electrodes. The coupling energy $E_m$ depends not only on $C_m$, but also on the total capacitance of the boxes: $E_m = 4e^2 C_m / (C_{\Sigma 1} C_{\Sigma 2} - C_m{}^2)$. Application of gate voltages allows us to control diagonal elements of the Hamiltonian (1). The circuit was fabricated to have the following relation between the characteristic energies: $E_{J1,2} \sim E_m < E_{c1,2}$. This ensures coherent superposition of the four charge states $|00\rangle$, $|10\rangle$, $|01\rangle$ and $|11\rangle$ around $n_{g1} = n_{g2} = 0.5$ while other charge states are separated by large energy gaps. The above condition justifies the use of a four-level approximation for the description of the system. In our notation $|n_1 n_2\rangle$ of the charge states used throughout the text, $n_1$ and $n_2$ refer to the number of excess Cooper pairs in the first and the second qubits, respectively.

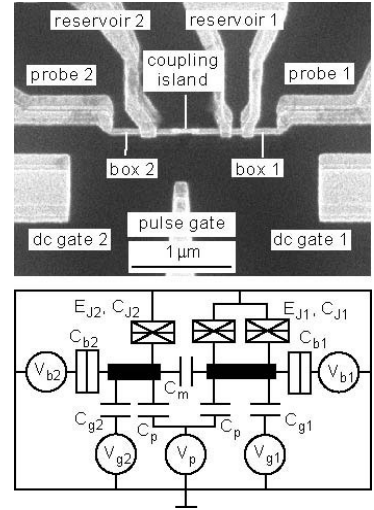

Figure 1. Two capacitively coupled charge qubits. **a**, Scanning electron micrograph of the sample. The qubits were fabricated by electron-beam lithography and three-angle evaporation of Al (light areas) on a SiN_x insulating layer (dark) (see ref.5 for fabrication details). Two qubits are coupled by an additional coupling island overlapping both Cooper-pair boxes. Although the coupling island has a finite tunnelling resistance ~ 10 MΩ to the boxes, we consider the coupling as purely capacitive represented by a single capacitor in the equivalent circuit because all the tunnelling processes are completely blocked. The estimated capacitance of the island to the ground is ~ 1 aF. **b**, Equivalent circuit of the device. The parameters obtained from the dc measurements are: $C_{J1} = 620$ aF, $C_{J2} = 460$ aF, $C_{b1} = 41$ aF, $C_{b2} = 50$ aF, $C_{g1} = 0.60$ aF, $C_{g2} = 0.61$ aF, $C_p \approx 1$ aF, $C_m = 34$ aF, and the corresponding energies are $E_{c1} = 484$ µeV (117 GHz in frequency units), $E_{c2} = 628$ µeV (152 GHz) and $E_m = 65$ µeV (15.7 GHz). Josephson coupling energies, $E_{J1} = 55$ µeV (13.4 GHz) and $E_{J2} = 38$ µeV (9.1 GHz), were determined from the single qubit measurements described later in the text. Probe junction tunnel resistance is equal to 31.6 MΩ (left) and 34.5 MΩ (right). Superconducting energy gap is 210 µeV. Black bars denote Cooper-pair boxes. Symbol ⊟ represents a tunnel junction without Josephson coupling, while ⊠ is a Josephson tunnel junction.

In the absence of Josephson coupling, the ground-state charging diagram $(n_1, n_2)$[14] (see Fig.2a) consists of hexagonal cells whose boundaries delimit two neighbouring charge states with degenerate electrostatic energies. For example, points R and L correspond to a degeneracy between the states $|00\rangle$ and $|10\rangle$ and the states $|00\rangle$ and $|01\rangle$ differing by one Cooper pair in the first and the second qubit, respectively. If we choose the dc gate charges $n_{g1}$ and $n_{g2}$ far from the boundaries but within the (0,0) cell, then because of large electrostatic energies we can assume that the system remains in the state $|00\rangle$. Since the pulse gate has equal coupling to each qubit, the application of a pulse shifts the state of the system on this diagram along the line tilted at 45 degrees indicated by arrows in Fig. 2a. The charging diagram remains valid for the small Josephson coupling except on the boundaries where charge states become superposed. When the system is driven non-adiabatically to the point R or L, it behaves like a single qubit oscillating between the degenerate states with a frequency $\omega_{1,2} = E_{J1,2}/\hbar$. Applying arrays of pulses and measuring oscillations of the probe currents $I_1$ and $I_2$, we can determine the Josephson energies of each qubit. The accuracy of the measured $E_{J1,2}$ is very high since the electrostatic coupling through $C_m$ has almost no effect on $\omega_{1,2}$ along the boundaries in the vicinity of R and L.

At the "co-resonance" point X ($n_{g1} = n_{g2} = 0.5$), the system has a double degeneracy, $E_{00} = E_{11}$, $E_{10} = E_{01}$, and the dynamics of the quantum evolution becomes more complex and reflects the coupling between the qubits. The cross-section of the energy bands through the point X is shown in Fig. 2b. Exactly at the co-resonance, all four charge states are mixed and the state of the system can be expressed in general as

$$|\psi(t)\rangle = c_1|00\rangle + c_2|10\rangle + c_3|01\rangle + c_4|11\rangle \qquad (2)$$

where $|c_i|$ ($i = 1, 2, 3, 4$) are the time dependent probability amplitudes obeying a normalisation condition $\sum_{i=1}^{4} |c_i|^2 = 1$.

Using the Hamiltonian (1) and initial conditions one can calculate the probabilities $|c_i|^2$ of each charge state. However, in our read-out scheme, we measure a probe current $I_{1,2}$ proportional to the probability $p_{1,2}(1)$ for each qubit to have a Cooper pair on it regardless of the state of the other qubit, i.e., $I_1 \propto p_1(1) \equiv |c_2|^2 + |c_4|^2$ and $I_2 \propto p_2(1) \equiv |c_3|^2 + |c_4|^2$. Assuming the initial state at $t = 0$ is $|00\rangle$, we can derive for an ideal rectangular pulse shape of a length $\Delta t$ the time evolution of these probabilities:

$$p_{1,2}(1) = (1/4)[2 - (1 - \chi_{1,2})\cos\{(\Omega + \varepsilon)\Delta t\} - (1 + \chi_{1,2})\cos\{(\Omega - \varepsilon)\Delta t\}] \qquad (3)$$

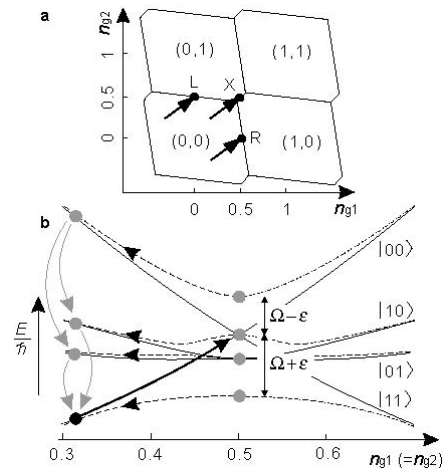

Figure 2. Pulse operation of the device. **a**, Schematics of the ground-state charging diagram of the coupled qubits as a function of the normalised gate charges $n_{g1}$ and $n_{g2}$. The number of Cooper pairs $n_1$ and $n_2$ in the neighbouring cells differs by one. The electrostatic energies $E_{n1n2}$ are degenerate at the boundaries. Points R and L correspond to energy degeneracy in the first and the second qubit, respectively. Point X is doubly degenerate: $E_{00} = E_{11}$ and $E_{10} = E_{01}$. Arrows show how pulses shift the system in the experiment. **b**, Energy diagram of the system along the line $n_{g1} = n_{g2}$ through the point X. Solid lines are the electrostatic energies of charge states $|00\rangle$, $|10\rangle$, $|01\rangle$ and $|11\rangle$. Dashed lines are eigenenergies of the Hamiltonian (1). Far from co-resonance (point X in **a**), the system stays in $|00\rangle$. After the pulse brings the system to the co-resonance (solid arrow), the system starts to evolve producing a superposed state $|\psi(t)\rangle = c_1|00\rangle + c_2|10\rangle + c_3|01\rangle + c_4|11\rangle$. The amplitudes $|c_i|$ ($i = 1, 2, 3, 4$) remain "frozen" after the pulse termination (dashed arrows) until the resulting state decays into the ground state. The decay process indicated by grey arrows contributes to the probe currents proportional to the probabilities (3).

where $\chi_{1,2} = (E^2{}_{J2,1} - E^2{}_{J1,2} + E_m{}^2/4)/(4\hbar^2 \Omega\varepsilon)$, $\Omega = ((E_{J1} + E_{J2})^2 + (E_m/2)^2)^{1/2}/2\hbar$, $\varepsilon = ((E_{J1,2} - E_{J2,1})^2 + (E_m/2)^2)^{1/2}/2\hbar$.

One can see that unlike a single qubit case, there are two frequencies present in the oscillation spectrum of the qubits: $\Omega + \varepsilon$ and $\Omega - \varepsilon$, both dependent on $E_{J1}$, $E_{J2}$ and $E_m$. We can identify these frequencies with the energy gaps in Fig.2b. Note that in the uncoupled situation ($E_m = 0$), $\Omega \pm \varepsilon = E_{J1,2}/\hbar$ and each qubit oscillates with its own frequency $\omega_{1,2}$. Let us stress, however, that the above consideration is valid only in the ideal case when the pulse has zero rise/fall time and the time evolution occurs exactly at the co-resonance point.

The idea of the experiment is shown schematically in Fig. 2b. From the state $|00\rangle$ (shown as a black dot), the pulse (solid arrow) brings the system to the co-resonance, and the system evolves for the pulse duration time $\Delta t$, producing a superposed state (2) indicated by grey circles. After the pulse termination (dashed arrows), the system remains in the superposed state until it decays (dotted arrows) in the ground state by emitting quasiparticles into the probe junctions biased at $V_{b1,2} \approx 600\ \mu eV$. To accumulate a signal, a pulse array ($\sim 3 \times 10^5$ pulses) was applied to the pulse gate. The repetition time between the pulses was 64 ns, long enough (in comparison to the quasiparticle relaxation time $\sim 10$ ns) to let the system decay through a Josephson-quasiparticle cycle[15] and give rise to a probe current proportional to $p_{1,2}$. The estimated amplitude of the applied pulses is $V_p \approx 30$ mV.

The results obtained in this way are presented in Fig. 3. First, by tuning $n_{g1}$ and $n_{g2}$, we do single qubit measurements by bringing the system to the point R or L and thus exciting autonomous oscillations in one of the qubits (Fig. 3a). The oscillations can be fitted to a cosine function with an exponential decay time of about 2.5 ns. The oscillations spectra (right panel of Fig. 3a) obtained by the Fourier transform contain one pronounced component at 13.4 GHz for the first qubit and at 9.1 GHz for the second one. We identify these values with $E_{J1}$ and $E_{J2}$. Judging from our previous experiments (see, e. g., ref. 13) we can conclude that these values are close to what we expect for the given fabrication parameters, i.e., overlap area and oxidation conditions. Then, by changing $n_{g1}$ and $n_{g2}$, the system is driven to the co-resonance and the induced quantum oscillations are traced using the same technique. The oscillation pattern becomes more complex (Fig.3b) and more frequency components appear in the spectrum. The observed spectral properties of the oscillations agree with the predictions of equation (3) in a sense that there are two peaks in the spectrum and the peak positions are close to the expected frequencies $\Omega + \varepsilon$ and $\Omega - \varepsilon$ for the parameters $E_{J1}$ = 13.4 GHz and $E_{J2}$ = 9.1 GHz measured in the single qubit experiments (Fig. 3a), and $E_m$ = 15.7 GHz estimated from the independent dc current-voltage characteristics measurements. The expected from equation (3) position of the $\Omega + \varepsilon$ and $\Omega - \varepsilon$ peaks is indicated by arrows and dotted lines. The decay time $\sim 0.6$ ns of the coupled oscillations is shorter compared to the case of independent oscillations as should be expected since an extra decoherence channel appears for each qubit after coupling it to its neighbour. The amplitudes of the spectral peaks, however, do not exactly agree with equation (3). We attribute this to the non-ideal pulse shape (finite rise/fall time $\sim 35$ ps), and the fact that a small shift of $n_{g1}$ and $n_{g2}$ off the co-resonance drastically changes oscillation pattern. Also, even far from the co-resonance, we still have a small contribution to the initial state from the other than |00> charge states distorting the oscillations. We have performed numerical simulation of the oscillation pattern taking into account a realistic pulse shape and not pure |00> initial condition assuming the system is exactly at the co-resonance. The resulting fits are shown in Fig. 3b as solid lines. We found that $E_m = 14.5$ GHz, close to the value estimated from the dc measurements, gives better agreement of the fit with the experimental data.

Finally, we checked the dependence of the oscillation frequencies on $E_{J1}$ controlled by a weak magnetic field (up to 20 Gs). The results are shown in Fig. 4. The plot contains the data from both qubits represented by open triangles (first qubit) and open circles (second qubit). Without coupling ($E_m = 0$), the single peaks in each qubit would follow dashed lines with an intersection at $E_{J1} = E_{J2}$. The introduced coupling modifies this dependence by creating a gap and shifting the frequencies to higher and lower values, the spacing between the two branches being equal to $E_m/2h$ when $E_{J1} = E_{J2}$. We compare the observed dependence with the prediction of equation (3) given by solid lines and find a remarkable agreement.

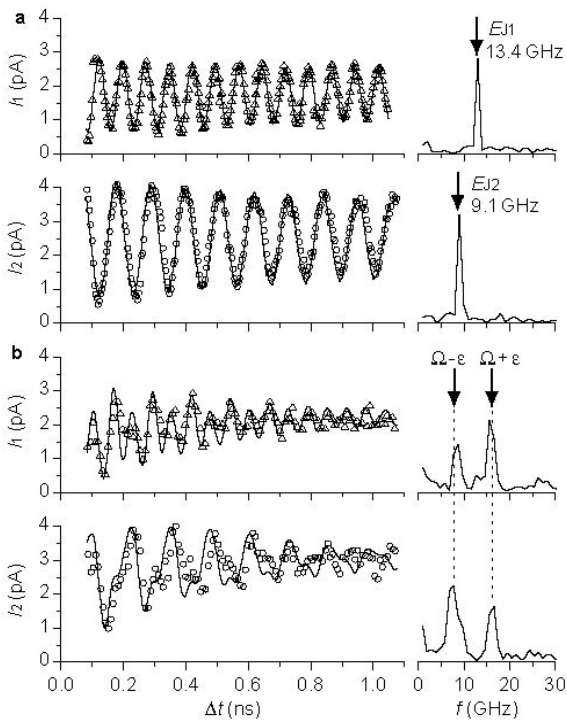

Figure 3. Quantum oscillations in qubits. **a**, Probe current oscillations in the first (top) and the second (bottom) qubit when the system is driven to the points R and L, respectively. Right panel shows corresponding spectra obtained by the Fourier transform. In both cases, the experimental data (open triangles and open dots) can be fitted to a cosine dependence (solid lines) with an exponential decay with 2.5 ns time constant. **b**, Probe current oscillations in the qubits at the co-resonance point X. Their spectra (right panel) contain two components. Arrows and dotted lines indicate the position of $\Omega + \varepsilon$, $\Omega - \varepsilon$ obtained from (3) using $E_{J1}$=13.4 GHz, $E_{J2}$=9.1 GHz measured in the single qubit experiments (Fig. 3a) and $E_m$=15.7 GHz estimated independently from dc measurements. Solid lines are fits obtained from numerical simulation with the parameters $E_{J1}$=13.4 GHz, $E_{J2}$=9.1 GHz and $E_m$=14.5 GHz. Finite pulse rise/fall time and not pure |00> initial condition were taken into account. The introduced exponential decay time is 0.6 ns.

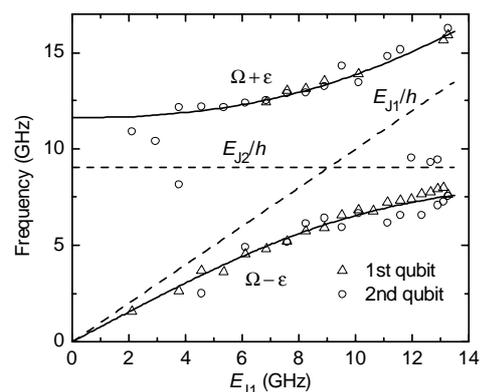

Figure 4. $E_{J1}$ dependence of the spectrum components obtained by the Fourier transform of the oscillations at the co-resonance. Open triangles: frequency components measured in the first qubit; open circles: frequency components measured in the second qubit; solid lines: dependence of $\Omega + \varepsilon$ and $\Omega - \varepsilon$ obtained from (3) using $E_{J2}$=9.1 GHz and $E_m$=14.5 GHz and varying $E_{J1}$ from zero up to its maximum value of 13.4 GHz; dashed lines: dependence of the oscillation frequencies of both qubits in the case of zero coupling ($E_m = 0$).

The observed quantum coherent dynamics of coupled qubits in the vicinity of the co-resonance (in particular, double-frequency structure of the probability oscillations in both qubits and frequency "repulsion'' at $E_{J1} \approx E_{J2}$ (see Figs. 3b and 4) indicates the two qubits become entangled during the course of coupled oscillations although direct measurement of the degree of entanglement was not possible. Simple calculation based on the standard expression for the entanglement of the pure states[16] show that with an ideal pulse shape and the |00> initial condition, the wavefunction (2) passes in its evolution through maximally entangled state in the case of equal Josephson energies. The amount of entanglement does not decrease significantly when realistic experimental conditions are taken into account that is confirmed by the numerical simulations. The relatively large observed oscillation amplitude (about 50% of the expected value) also suggests the existence of entangled states even in our multi-pulse averaged experiment.

In conclusion, we were able to manipulate quantum states of two coupled Josephson charge qubits using a common pulse gate and observed time-domain oscillations with a clear evidence for interaction between the qubits. With these results, construction of a solid-state quantum logic gate could be envisaged.

* on leave from P. N. Lebedev Physical Institute, Moscow 117924, Russia

We thank B. L. Altshuler, X. D. Hu, H. Im, S. Ishizaka, F. Nori, T. Sakamoto and J. Q. You for fruitful discussions. D.V.A. was supported by the AFORS and by the NSA and ARDA under the ARO contract.

Correspondence should be addressed to J. S. T. (e-mail: tsai@frl.cl.nec.co.jp).